\documentclass[preprint,apl]{revtex4-1}
\usepackage{epsfig}
\usepackage{color}
\usepackage{newlfont}
\usepackage{verbatim}
\usepackage{amssymb,amsmath}

\begin{document}

\title{Characterization and In-situ Monitoring of Sub-stoichiometric Adjustable $ \textrm T_\textrm C $ Titanium Nitride Growth}
\author{ Michael R. Vissers }
\email {michael.vissers@nist.gov}
 \author{Jiansong Gao}
 \author{ Jeffrey S.  Kline}
 \author{ Martin Sandberg}
 \author{ Martin P. Weides}
 \altaffiliation{Current address: Karlsruhe Institute of Technology, 76131 Karlsruhe, Germany}
  \author{David S. Wisbey}
  \altaffiliation{Current address: Department of Physics, Saint Louis University, 3450 Lindell Blvd. Saint Louis, MO 63103, USA}
 \author{David P. Pappas}
\email{david.pappas@nist.gov}
\affiliation{National Institute of Standards and Technology, \\ 325 Broadway, Boulder, CO, 80305}

\date{\today}

\begin{abstract}

The structural and electrical properties of Ti-N films deposited by reactive sputtering depend on their growth parameters, in particular the Ar:N$_2$ gas ratio. We show that the nitrogen percentage changes the crystallographic phase of the film progressively from pure $\alpha$-Ti, through an $\alpha$-Ti phase with interstitial nitrogen, to stoichiometric Ti$_2$N, and through a substoichiometric TiN$_X$ to stoichiometric TiN.  These changes also affect the superconducting transition temperature, $T_C$, allowing, the superconducting properties to be tailored for specific applications. After decreasing from a $T_C$ of 0.4 K for pure Ti down to below 50 mK at the Ti$_2$N point,  the $T_C$ then increases rapidly up to nearly 5 K over a narrow range of nitrogen incorporation.  This very sharp increase of $T_C$ makes it difficult to control the properties of the film from wafer-to-wafer as well as across a given wafer to within acceptable margins for device fabrication.  Here we show that the nitrogen composition and hence the superconductive properties are related to, and can be determined by, spectroscopic ellipsometry. Therefore, this technique may be used for process control and wafer screening prior to investing time in processing devices.

\noindent Contribution of U.S. government, not subject to copyright.
\end{abstract}


\maketitle 

\section{Introduction}
Microwave kinetic inductance detectors, MKIDs, have significant promise in astronomical detector applications \cite{Day-MKID}.  A generic MKID consists of a microwave resonator that is coupled to a feedline, leading to a dip in the transmission at the resonator's resonant frequency. One of the most important aspects of these devices is that they can be frequency multiplexed by fabricating many resonators with slightly different resonant frequencies, and thus they all can be coupled to a common readout bus. In this case, each resonator rings at a frequency $\textrm{f}_\textrm{r}=1/(2\pi\sqrt{\textrm{LC}})$, where the L and C are the device's inductance and capacitance, respectively.  While in most instances the inductance and capacitance are defined overwhelmingly by geometry, for superconductors the inductance has an additional term due to the finite inertia of the Cooper pairs, i.e., the kinetic inductance.  From Mattis-Bardeen theory, the kinetic inductance can be approximated\cite{Jonas-review} by $\hbar \rho / (\textrm{t}\pi\Delta) $, where $\rho$ is the film's resistivity, $\textrm{t}$ is the thickness and $\Delta$ is the film's superconducting gap.   When an incident photon is absorbed by the superconducting resonator, the photon breaks Cooper pairs creating excess quasiparticles.  These quasiparticles increase the kinetic inductance and shift the resonance to a lower frequency.  Hence, the material's intrinsic kinetic inductance and superconducting transition temperature, $T_C$, are critical parameters to increase the sensitivity and ensure that photons in the desired frequency range f have the requisite energy to be absorbed and create quaisparticles, i.e. $\textrm{f} > 2\Delta/\textrm{h}$ where $2\Delta$ is the Cooper-pair binding energy and h is Planck's constant.  More specifically, astronomical applications such as the cosmic microwave background (CMB) search desire detection of 90 GHz light, necessitating a $T_C\le$ 1 K.  

In addition to matching this stringent $T_C$ requirement, superconducting materials for MKIDs should exhibit other properties such as a high kinetic inductance, low loss (i.e. high quality factor) and low frequency noise. Recent advances in the use of titanium nitride (TiN) have shown that it meets these criteria, \cite{JPL-TiN, Vissers-TiN, Jonas-review} resulting in greater sensitivity than elemental superconductors such as Nb or Al.  Just as importantly, TiN also has a tunable $T_C$. While stoichiometric TiN has a $T_C$ above 4.5 K, substoichiometric TiN$_x$ (x $<$ 1) exhibits a $T_C$ that can be considerably lower.  The nitrogen percentage also alters the film crystallography and resistivity, affecting the film's kinetic inductance and the resulting device performance.  In this paper we investigate how the structure and $T_C$ of the Ti-N compounds grown by reactive sputtering change by varying the ratio of the Ar-N$_2$ gas inside the sputtering chamber during deposition. We also apply spectroscopic ellipsometry to these films and show that the film's measured ellipsometric properties are correlated to the final $T_C$.  

The binary phase diagram of titanium and nitrogen shows that several different compounds are stable as the Ti:N atomic ratio is altered at our $ \le 500 ^{\circ} C$ growth temperatures\cite{Wriedt}.  For the lowest N$_2$ percentages an $\alpha$-Ti phase is initially formed, where N is incorporated interstitially in the Ti lattice. This is followed by a narrow band near 33 $\%$ atomic N fraction of the compound Ti$_2$N.  At higher nitrogen fractions, TiN$_x$ is stable in a broad range from a substoichiometric x=0.6 to an overstoichiometric x=1.2 \cite{Greene}.  Previously\cite{JPL-TiN}, varying the nitrogen fraction in the gas mixture used to reactively sputter Ti-N compounds has been shown to alter the measured $T_C$ from 0 K to over 4.5 K.  

The composition of the deposited Ti-N compounds from reactive sputtering can be altered by changing the effective flux of Ar:N atoms impinging on and reacting with the Ti sputter target.  During sputtering, there is competition between two different processes: lighter N ions react with Ti on the surface to form Ti-N, and inert Ar ions sputter the Ti atoms and Ti-N molecules from the target \cite{Petrov}.  The amount of freshly exposed Ti at the target from the Ar sputtering determines the amount of available sites for nitrogen reaction.  If the nitrogen flow exceeds the rate at which the Ti is exposed, the surface of the target will be completely nitrided, i.e. poisoned, and only TiN will be sputtered.  Conversely, if the Ar removes the Ti faster than it can be nitrided, then both Ti and TiN will be sputtered onto the substrate with a controllable ratio.  Assuming that all the reactive nitrogen is consumed at the target, this ratio of deposited Ti:TiN determines the composition of the film.   However, as all the gases in our chamber are delivered centrally in the chamber instead of directly at the gun, additional nitridization of non-fully bonded Ti is also seen at the substrate, as is shown below.  This dependence indicates that the TiN$_x$ is not fully in a stable phase when deposited, and is sensitive to additional nitridization on the substrate.  Although the local change in the TiN$_x$ nitrogen percentage is minor, since the $T_C$ of the film is so sensitive to composition at the steepest point of the curve a substantial variation in the measured $T_C$ across the wafer has been observed\cite{JOLT,Gao_SubStoichometricTiN_arXiv}.  
\section{Experimental Details}
The TiN films in this work are sputter deposited onto 3" (100) oriented Si substrates in a chamber with a base pressure of $1\times10^{-9}$ torr at room temperature.  A three inch diameter, 99.995 \% pure Ti target is sputtered in a reactive atmosphere of Ar and N$_2$, depositing TiN onto the rotating substrate.  The Ti target is initially pre-sputtered in an atmosphere of pure Ar to remove any TiN or other contamination from the target surface. The target preparation is important because the growth process of TiN is hysteretic, i.e. the deposition rate depends upon whether the target starts in the clean-Ti or fully poisoned TiN limit.  During the depositions, the total pressure was kept fixed by throttling a variable position gate valve at $5 \pm 0.1$ mtorr. The Ar flow was maintained at $15 \pm 0.1$ sccm for all of the depositions, and the N$_2$ flow was modified in order to obtain the various Ar:N flux ratios at the target.    After the N$_2$ flow is started, but before all the shutters were opened to begin the growth, the plasma was allowed to settle for 6 minutes to allow for N$_2$ consumption by the Ti target, and subsequent sputtering of Ti and TiN, to come to equilibrium. We then deposited nominally 35 nm of TiN$_x$ on each sample for the various Ar:N flux ratios.   Before and after each growth, \emph{in-situ} spectroscopic ellipsometry data were taken at wavelengths from 200 nm to 1000 nm.

After the growth is completed, \emph{ex situ} position dependent ellipsometry is also performed, and the wafer is then fabricated into unpatterened 5x1 mm sheet resistance and co-planar waveguide resonators (CPW) test structures, diced into chips, and cooled for $T_C$ testing in an adiabatic demagnetization refrigerator (ADR). The ADR has a base temperature of 50 mK and allows for a continuously variable temperature from the base temperature to 10 K.  The $T_C$ was measured using a DC current source and voltage meter. The measured $T_C$ at the center of the different wafers as a function of N$_2$ gas flow is shown in Figure 1 and is of a similar shape as that previously reported\cite{JPL-TiN}.  
\section{Results and Discussion}
While the $T_C$ varies slowly in both the high and low limits of the N$_2$ flow, the measured $T_C$ sharply changes from 0.1 K to 3 K for small changes in nitrogen flow near 2 sccm.  The target $T_C$ of 1 K is located on this steep slope.  In this region, changing the Ar-N$_2$ ratio by less than $0.5 \%$  leads to the $T_C$ changing by more than 1 K. Moreover, as the $T_C$ is so sensitive to the nitrogen percentage in this region, we begin to see substantial variations in the film properties across the 3" wafer, likely from additional nitridization of unbonded Ti at the substrate.  As shown in Figure 2(b), the $T_C$ varies by more than $25 \%$ from the center of the wafer to the edge despite the deposition rate varying by less than 1 $\%$ across the wafer. Furthermore, neither the $T_C$ nor the ellipsometric properties are strong functions of the thickness.  Films sputtered twice as thick or 50 \% thinner with the same deposition conditions have a $T_C$ range that is almost identical.  While the thinner films are no longer optically opaque, and thus have different measured optical properties, the thicker films have ellipsometric measurements that are indistinguishable from the 35 nm films with the same $T_C$.   The rotating substrate decreases the variation in $T_C$; static substrates (not shown here) generally have more than twice the anisotropy in $T_c$.


X-ray diffraction (XRD) was also performed on these samples  by use of a $\theta$:$2\theta$  instrument.   The changing crytallographic regions as a function of nitrogen content and in relation to the measured $T_C$ are also shown in Figure 1. Similar to what has been seen previously in reactively sputtered TiN films,\cite{Sundgren,JOLT} the films with higher nitrogen fraction and $T_C$ are seen to have a mixture of the TiN (111) and (200) orientations.   As the nitrogen flow is reduced, the mixture evolves to a wholly (111) TiN orientation, with $T_C$ dropping through 1 K and down to 0.4 K.  However, for lower nitrogen percentages, the $T_C$ drops further and greater incorporation of the Ti$_2$N phase is seen.  When the film appears to be fully Ti$_2$N, the $T_C$ is below the 50 mK lower limit of our measurement apparatus.    Stoichiometric Ti$_2$N is metallic, is known to be non-superconducting to below 1.2 K\cite{Spengler}, and its presence suppreses $T_C$ in Ti-N compounds\cite{Radhakrishnan}. To the authors' knowledge, no superconducting transition has ever been measured for Ti$_2$N.  This implies that the mono-phase Ti$_2$N films with a $T_C <$ 0.05 K are not unreasonable.  For films with even less N$_2$, $T_C$ becomes measureable again, but as revealed by X-ray diffraction the material is now $\alpha$-Ti phase.  Pure Ti films were also grown with a $T_C$ of 0.4 K.


The steep profile of $T_C$ vs N$_2$ flow is repeatable over day to day timescales, but drifts slowly over months, likely due to target erosion, and offsets occur during chamber maintenance or after large excursions of the mass flow controller (MFC). The N$_2$ MFC has a nominal repeatability of 0.25 sccm, but in typical applications with just small changes in flow rates between growths we measure an uncertainty of only 0.05 sccm.  If we compare this uncertainty to the slope of the line in Figure 1, this implies an uncertainly of 0.5 K in the $T_C$ of the film.  While the $T_C$ vs N$_2$ flow curve can be re-calibrated again using many growths and cooldowns, an accurate \emph{in situ} or pre-cooldown probe of the TiN composition would greatly reduce the period of time needed to refine the TiN growth and also provides a pre-fabrication check of the viability of the film for the designed detectors. It has been shown previously that \emph{in-situ} ellipsometry is senstive to the composition of Ti-N compounds both near the stoichiometric limit \cite{Logo-JAP, Logo-SCT, adachi} as well as for lower nitrogen fractions \cite{Bendavid,Ahn}.  Ellipsometry has also shown to be senstive to the N incorporation in NbTiN thin films\cite{Cecil}.

Ellipsometry is a non-destructive optical technique that illuminates a sample with an oblique beam of polarized light. The change in the polarization state of the reflected beam is measured, commonly expressed as $\rho = r_{\textrm{p}} / r_{\textrm{s}}=\tan(\Psi)e^{i\Delta}$ where $r_{\textrm{p}}$ and $r_{\textrm{s}}$ are the reflectivities of the $\textrm{p}$ and $\textrm{s}$ polarized light respectively.  In spectroscopic ellipsometry, these values are acquired as a function of wavelength, greatly increasing the data set and permitting the determination of multiple physical properties of the sample.  Here, the ellipsometer is mounted onto the vacuum chamber and the light is shone through low stress windows onto the sample at an incident angle of $\approx$ 70 degrees. These \emph{in situ} ellipsometric data can be used to optimize growth parameters both during and after the growth.  

Figure 3 shows that that the measured ellipsometric properties of TiN films are very sensitive to the film composition.  As the nitrogen concentration is reduced and the $T_C$ changes, the measured ellipsometric property $\Delta$ monotonically changes as seen in Fig 3(a).  In addition to the measured ellipsometric properties, $\Psi$ and $\Delta$, the same information can also be expressed in terms of the pseudo-dielectric function $(<n>+i<k>)^2=<\epsilon_1 +i\epsilon_{2}> =\sin^{2}\phi + (\frac{1-\rho}{1+\rho})^2\tan^2\phi\sin^{2}\phi $ where $\phi$ is the angle of incidence. The measured data from Figure 3(a) can then be re-plotted in Figure  3(b), the real part of the dielectric function, $<\epsilon_1>$, vs photon energy $E=\frac{hc}{\lambda}$, where $\lambda$ is the photon wavelength.  While the difference in $<\epsilon_1>$ between the films with different $T_C$'s is not as striking as those seen in the measured $\Delta$, the changes do illustrate how the compositional changes in the various films are manifested.  The photon energy at which $\epsilon_1$, the real part of the dielectric function, passes through zero is generally referred to the unscreened plasma energy, $\omega_p=\sqrt{\frac{Ne^2}{\epsilon_0 m^*} }$ where $N$ is the number of carriers and $m^*$ is their effective mass.   Figure 4 shows the zero crossing of $<\epsilon_1>$ for the different TiN$_x$ films as a function of $T_C$.  The pure Ti film's zero crossing is not within our observable energy range. As the nitrogen is reduced, the $T_C$ drops and the energy of the zero-crossing increases. This implies that there is an increased number of carriers as the films become more metallic.  This result is consistent with the XRD and phase diagram results as Ti$_2$N \cite{Igasaki} is also a metal, with a higher conductivity and carrier density than TiN.   The zero crossing of $<\epsilon_1>$ is composition dependent and is a sensitive indication of the $T_C$ of the Ti-N film to within 0.5 K. The inset of Figure 4 shows the value of the measured ellipsometric quantity $\Delta$ at 300 nm photon wavelength.  While the change in $\Delta$ does not directly correspond to any physical property, it provides a more sensitive lookup table for the film $T_C$.  

When mounted to the chamber in the \emph{in situ} configuration the optical axis is fixed, but the ellipsometer can also be mounted off the chamber with a moveable stage and ellipsometric measurements can be made as a function of position.   Figure 2 (a) shows how the energy of the zero crossing of the real part of the dielectric function $\epsilon_1$ changes across the wafer in a manner that resembles the $T_C$ measured in Figure 2 (b).  While the ellipsometer is sensitive to the no longer pristine surface, and the measurements cannot be directly mapped to the previous \emph{in-situ} $T_C$ measurements, this measurement shows how the ellipsometer can be used to measure the homogeneity of the TiN films, and predict the relative variation in $T_C$.  
\section{Conclusion}
In conclusion, titanium nitride films were grown by reactive sputtering across a range of Ar to N$_2$ ratios at room temperature.  By varying the gas ratio, the composition and structure of the grown film is varied.  The crystallographic structure changes from TiN to Ti$_2$N to $\alpha$-Ti  as the nitrogen is reduced and the $T_C$ reduces as well from above $ 4.5$ K to less than $0.05$ K when the film is fully in the Ti$_2$N state before slightly rising again at the lowest N$_2$ flows.  Optical measurements are sensitive to these compositional changes as well. \emph{In situ} ellipsometric data clearly show the dependence on N$_2$ flow and can be used to predict and subsequently tune the $T_C$ of the film without extensive processing and cryogenic testing.  \emph{Ex situ} position dependent measurements also illustrate the ellipsometer's sensitivity, and provide a method to measure the film's homogeneity.







\begin{acknowledgements}  We acknowledge support for this work from DARPA, the Keck Institute for Space Studies, the NIST Quantum Initiative, and NASA under Contract No. NNH11AR83I. The authors thank Jonas Zmuidzinas and Henry Leduc for helpful discussions and insights. This work is a contribution of U.S. Government, not subject to copyright. The views and conclusions contained in this document are those of the authors and should not be interpreted as representing the official policies, either expressly or implied, of the U.S. government.

\end{acknowledgements}

\newpage

\begin{figure}[ht]\centerline{\includegraphics[width = 8.5 cm]{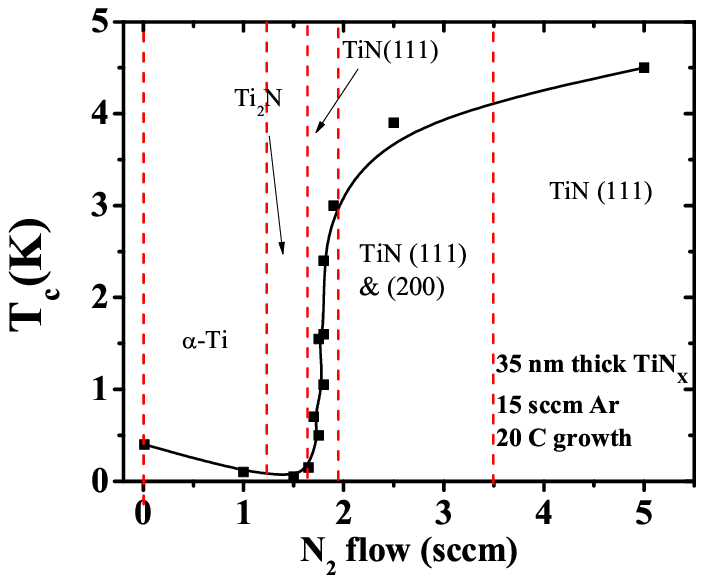}}
\caption{ Superconducting transition temperature $T_C$ of TiN films grown with different nitrogen flows. $T_C$ changes rapidly in the region of interest around 1 K.  Also indicated on the figure are the different crystallographic regimes as measured by $\theta:2\theta$ x-ray diffraction measurements.  }
\label{Fig1}
\end{figure}

\begin{figure}[ht]\centerline{\includegraphics[width = 17 cm]{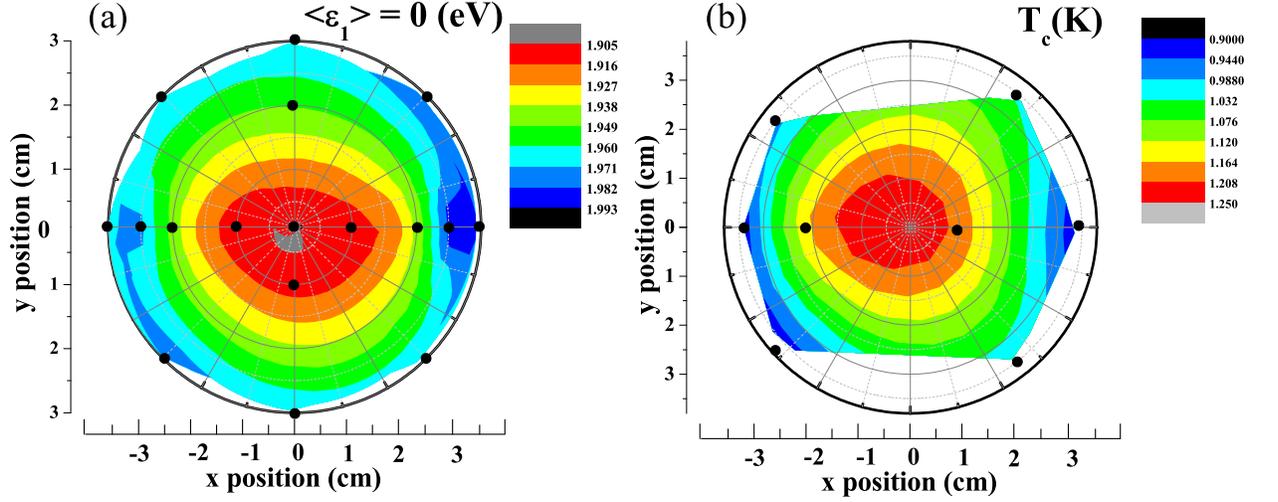}}
\caption{ (a) A position dependent contour plot of the  photon energy at which the real part of the TiN dielectric function, $<\epsilon_1>=0$ as measured by the ellipsometer after being removed from the deposition tool. The contour plot is an interpolation of 17 measured locations. (b) The measured $T_C$ across the TiN wafer.  The plot is an interpolation from 8 $T_C$ measurements shown as points in the plot.  The similarity in the two wafermaps illustrates the sensitivity of the ellipsometer to the TiN composition and resulting film $T_c$.    }
\label{Fig5}
\end{figure}


\begin{figure}[ht]\centerline{\includegraphics[width = 17 cm]{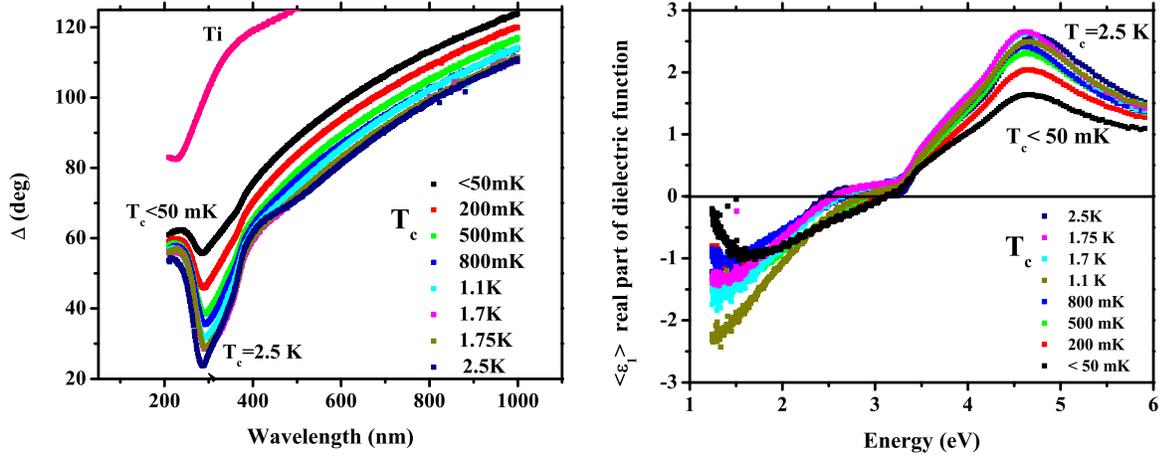}}
\caption{ (a) The measured ellipsometric parameter $\Delta$ vs wavelength for TiN films with different growth parameters along the sharp cliff in $T_C$.  The ellipsometer is sensitive to the difference in Ti-N compound composition. (b) The real part of the dielectric function, $<\epsilon_1>$, versus photon energy of the same films calculated from the measured ellipsometric parameters $\Psi$ and $\Delta$. The zero crossing of the $<\epsilon_1>$ is plotted in Figure 4. 
}
\label{Fig3}
\end{figure}

\begin{figure}[ht]\centerline{\includegraphics[width = 8.5 cm]{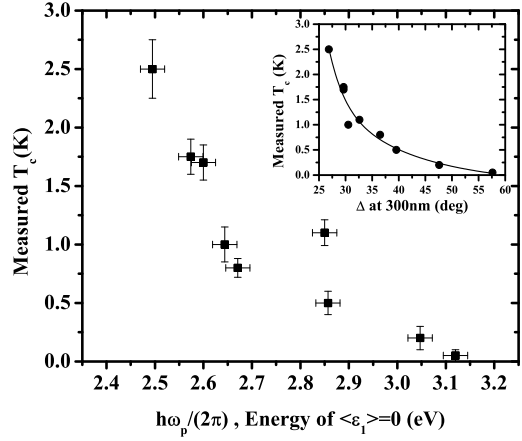}}
\caption{$T_C$ vs the photon energy where $<\epsilon_{1}>=0$.  This energy determines the unscreened plasma frequency $w_p=\sqrt{\frac{Ne^2}{m^*\epsilon_0}}$.    As less nitrogen is incorporated into the film in the area around the steep cliff in $T_C$, $T_C$ decreases but the unscreened plasma energy $\hbar\omega_p$ and hence the number of carriers increases as the films become more metallic.  A pure Ti film has an energy above the 6 eV measurement maximum in our setup.  The y-error bars are the variation in the $T_C$ across each film.  Inset:  The measured ellipsometric $\Delta$ at 300 nm photon wavelength vs Tc.  While $\Delta$ is not a physical property of the film, it provides a better lookup table for the TiN $T_C$.     }
\label{Fig4}
\end{figure}


\end{document}